\RequirePackage{fix-cm}
\documentclass[twocolumn,epjc3]{svjour3}
\smartqed  
\RequirePackage{graphicx}
\usepackage{url}
\usepackage{subfigure}
\usepackage{graphicx,color,dcolumn}
\usepackage{url}
\usepackage{amsmath}
\RequirePackage{color}
\usepackage{color}

\usepackage{titlesec}
\titlespacing*{\subsection}{0pt}{0.5em}{0.5em}

\begin{document}

\title{Renormalizable Cosmology Based on Gauss-Bonnet Theory with Torsion } \subtitle{}
\author{\makebox[0pt][l]{Zi Rui Hu\thanksref{addr1,addr2,addr3}, Zhi Fu Gao$^{\dag}$\thanksref{e1,addr1,addr4}, Hao Xuan Sun\thanksref{addr3}, Ci Xing Chen$^{\dag}$\thanksref{e2,addr3}, Xiao-Feng Yang\thanksref{addr1}}}

\thankstext{e1}{e-mail: zhifugao@xao.ac.cn}
\thankstext{e2}{e-mail: daleccx@ustc.edu.cn}
\institute{Xinjiang Astronomical Observatory, CAS, 150, Science 1-Street, Urumqi, Xinjiang, 830011, China.
\label{addr1}
\and Department of Astronomy, University of Sciences and Technology of China, CAS, Hefei,Anhui, 230026, China.
\label{addr2}
\and Brown University, 69 Brown St Box 1822, Providence, RI 02912, USA.
\label{addr3}
\and Key Laboratory of Radio Astronomy, CAS, Nanjing, Jiangsu, 210008, China
\label{addr4}
}
\date{Received: date / Accepted: date}

\maketitle
\begin{abstract}
  This paper focuses on renormalizable cosmology based on the Gauss-Bonnet theory with torsion. 
Within the framework of renormalizable quantum field theory, we study the matter field containing the 
Gauss-Bonnet correction term. By modifying the gauge model, which includes a charged scalar field and 
two families of fermions, and introducing torsion and Gauss-Bonnet corrections, we analyze the field 
equations in the supersymmetric hybrid inflation model, investigate the properties of torsion in a flat 
universe, the characteristics of the energy-momentum tensor, and the sensitivity of the theoretical model 
to correction parameters. This study shows that the Gauss-Bonnet correction term can directly affect 
the Hubble constant through the Einstein equations, providing a new observational perspective. This paper 
proposes a cosmological model that can be adjusted according to different Gauss-Bonnet correction models, 
providing a theoretical reference for future observational verification of various Gauss-Bonnet models.

\end{abstract}

\section{Introduction}
\label{sec:1}
\subsection{\textbf{Overiew}}
So far, fundamental physical theories face many challenging problems. For example, we 
have not yet fully understood the characteristics of gravity on a cosmic scale, leading 
to the proposal of alternative theories of gravity\,\cite{Carroll2004,Linder2010}.
Explaining the accelerated expansion of the universe from fundamental physics is a 
significant challenge, and observations of supernovae suggest the existence of a positive 
cosmological constant, but there is a lack of robust theoretical representation for the 
inflation. Addressing these problems often requires modifications to widely accepted cosmological 
theories to construct an extended standard cosmological model that includes quantized 
gravity treatments.\\
As a natural extension of Einstein's general relativity, the Einstein-Cartan (EC) theory of gravity 
considers the spin properties of matter and describes their effects on the geometric structure of 
spacetime, known as Riemann-Cartan spacetime, characterized by the non-trivial curvature and torsion 
(for example, see~\cite{ObukhovandKorotky1987,GarciadeAndrade2013,GarciadeAndrade2022,GarciadeAndrade2024}). However, background torsion breaks the weak equivalence principle
\,\cite{Ciufolini2019,Touboul2022} and violates local Lorentz invariance\,\cite{Kostelecky2008}. 
Additionally, there is no experimental or observational evidence supporting the unique predictions 
of the EC theory, mainly because the theory only deviates from classical general relativity under 
extremely high energy densities.\\
The Standard Model(SM) is a comprehensive theory in particle physics that describes the fundamental 
particles and their interactions, excluding gravity. String theory is an attempt to quantize gravity 
(for example, see~\cite{Schwarz1982,Polchinski1994,Giveon1994,AshtekarBianchi2024}). 
Although it primarily exists as a mathematical model in the high-energy framework due to its stringent 
observational requirements and predictive challenges, string theory possesses interesting mathematical 
properties and naturally incorporates many features of the SM, such as (1)non-Abelian groups: 
In the SM, the non-Abelian groups include the $\mathrm{SU}(2)$ and $\mathrm{SU}(3)$ groups, which are well represented 
in string theory, and (2)Chiral fermions: string theory can naturally include chiral fermions, which are 
crucial for describing the fermions in the SM\,\cite{DeBoer1994,Mambrini2006,Cheng2024}. 
These characteristics make string theory a strong candidate for a unified description of particle physics 
and gravity. 

String theory, in its low-energy limit, can indeed provide potential insights for observations, 
as it reduces to effective field theories that can be compared with experimental data\,\cite{Grana2006,Cappelli2012,SinhaZahed2021}. It does not necessarily require additional fundamental fields 
beyond those already present in general relativity and quantum field theory. Also, string theory 
avoids Ostrogradsky instability, which has been proposed as an explanation as towhy no differential 
equations of higher order than two appear to describe physical phenomena\cite{HayatoTeruaki2015}. The 
coupling of string theory with other fields can lead to novel dynamic effects, which could be observable 
and provide feasibility for observational studies.

Another influential candidate theory is the Loop Quantum Gravity (LQG). As well as string theory, the 
LQG is one of the most successful quantum gravity theories to date. It approaches the quantization of 
gravity by analogizing general relativity to gauge field theory \cite{AshtekarBianchi2024}. In this method, 
the Ashtekar-Barbero connection replaces the metric tensor as the fundamental generalized coordinate variable, 
and the quantization is achieved using loop integrals of the connection. Although this paper does not directly 
discuss this approach, it shares similarities with LQG as it rewrites the standard field theory using 
torsion as a variable independent of spacetime and explores their interaction with matter fields\cite{Raby2006}.

As a low-energy limit of effective string theory, the Gauss-Bonnet(GB) gravity, which 
belongs to the Lovelock gravity family\cite{Lovelock1971}, has become a hot topic in the field of 
cosmology \cite{Raby2006,Anson2019}. It is a ghost-free extension of Einstein's gravity, 
with the left-hand side of its field equations being a symmetric, conserved tensor that does not contain derivatives 
of the metric higher than second order. Recently, the GB theory has garnered considerable attention and 
has developed into various models and forms. For example, Glevan and Lin \cite{GlevanLin2020} proposed 
a universally covariant modified gravity theory that propagates only massless gravitons, making an important 
category Einstein-Gauss-Bonnet (EGB) gravity theory\cite{Bhattacharjee2017,BlazquezSalcedo2020} 
possible in four-dimensional spacetime. The EGB theory has decisive impacts on neutron stars, gravity, 
black holes\,\cite{Gao2016,Gao2017,Wen2020,Wen2021,Gao2019,Gao2020,Feng2022}, and cosmology\,\cite{Feng2024,LinYang2009,YangLin2019} by the non-minimally couplings. Among many gravity 
theories, those with higher curvature corrections have \\received considerable attention in recent 
years\,\cite{Meissner2004,CanateBreton2019,Hod2019}. 
\subsection{\textbf{ Torsion and Gauss-Bonnet Invariant $\mathcal{G}$ }}
In the realm of general relativity, torsion represents the antisymmetric component of the affine connection, 
describing the twisting or rotational geometric phenomena of a manifold in curved space, 
and illustrating how vectors evolve in curved spacetime. The classical GB theory involves the integration of 
curvature and boundary curvature. However, when “torsion” is introduced, this theory undergoes 
some extensions and changes.  As a unique feature of spacetime, torsion has specific advantages in dynamical analysis 
and can, to some extent, serve as an alternative gravitational theory\,\cite{Bahamonde2023a,Bahamonde2023b}.

Although torsion may not directly capture the dynamics and characteristics of fields, it proves advantageous in 
tackling challenges beyond the reach of conventional classical theories, such as those posed by black holes
or the early universeis. For instance, in the Einstein-Cartan theory, the incorporation of non-zero torsion into 
the equations of motion elevates this previously neglected spacetime feature to a significant role in extreme 
gravitational fields\,\cite{Meissner2004,CanateBreton2019}. This integration of torsion enhances the 
theory's ability to elucidate phenomena in strongly curved regions, where traditional general relativity may fall short.
 
The GB invariant is an important concept in differential geometry and theoretical physics. It is often expressed 
in the context of higher-dimensional gravity theories, such as Lovelock gravity.  As a stringy correction, the 
term of the GB invariant $\mathcal{G}$ is definede as:
\begin{equation}
\mathcal{G}=R_{\mu\nu\rho\sigma} R^{\mu\nu\rho\sigma} - 4 R_{\mu\nu} R^{\mu\nu} + R^2,
\label{1}
\end{equation}
where $R_{\mu\nu\rho\sigma}$ is the Riemann curvature tensor, $R_{\mu\nu}$ is the Ricci curvature 
tensor, and $R$ is the Ricci scalar\,\cite{Fomin2020,Fernandes2022a}. It arises in the low-energy effective 
action for the heterotic strings, and also appears in the second order of Lovelock gravity theory. When this 
invariant term interacts with matter fields, it induces non-trivial dynamic effects, significantly 
influencing the evolution of these fields. 

The GB term with torsion can serve as a higher-order correction term, addressing the limitations of 
classical general relativity in strong gravitational fields. The GB theory with torsion plays several key 
roles in renormalizable cosmology, primarily manifested as follows\,\cite{Hod2019}:
\begin{enumerate}
\item By introducing quadratic terms of Ricci curvature, the GB theory improves the 
 ultraviolet divergence behavior of the theory, making it more stable at high energy scales.

\item Although the GB invariant with torsion may violate causality, it 
  can maintain causality in four-dimensional spacetime through appropriate redefinition of the 
 coupling constant, such as  $\alpha_2 \to \frac{\alpha_2}{D-4}$.

\item In four-dimensional spacetime, the GB invariant with torsion acts 
 as a topological term, not requiring the introduction of new degrees of freedom but making significant 
 contributions to loop diagram behavior during renormalization.
\end{enumerate}
\subsection{\textbf{Motivation and Innovation}}
Renormalization is a mathematical method that was originally developed in quantum field theory. Its primary goal is to eliminate infinities from calculations by adjusting the theory's parameters. By introducing renormalization factors, divergent terms are transformed into finite quantities, ensuring the theory remains self-consistent across different scales. This process allows us to describe the formation of large-scale structures in the universe as emerging from the behavior of fundamental particles.\\
\indent In this work, we focus on renormalizable cosmology based on Gauss-Bonnet (GB) theory with torsion. Unlike previous studies, we introduce the GB term into a gauge model, allowing us to explore the effects of the GB term on the matter field without altering the characteristics of spacetime. Our research uniquely combines the supersymmetric hybrid inflation model with torsion as an independent spacetime variable and examines its interaction with the matter field, leading to modifications in spacetime theory. The study reveals that the GB term can directly influence the Hubble constant, $H$, through Einstein’s equations, providing a novel observational perspective, thereby making this research both theoretically significant and practically relevant.\\
\indent This work proposes a cosmological model adjustable to various formulations of GB theory. Furthermore, our model facilitates the derivation of other related theories through field equations, such as the dynamic behavior of cosmology, the inflationary process of the early universe, and various phenomena associated with the distribution of matter and energy. Throughout this paper, we adopt natural units where $c = \hbar = 1$, and Newton’s gravitational constant is given by $8\pi G = m_P^{-2}$, where $m_P = 2.4 \times 10^{18}\,\text{GeV}$ is the reduced Planck mass.\\
\indent The structure of the paper is as follows: Section 2 reviews classical torsion theory, constructs the interaction between torsion and fields, and derives the renormalized action. Section 3 presents new Einstein equations with Gauss-Bonnet corrections by calculating the energy-momentum tensor for the matter field with torsion. Section 4 integrates the GB invariant into the supersymmetric hybrid inflation model and applies the GB correction to the field equations. Finally, Section 5 summarizes our findings and discusses future research directions.

\section{Renormalizable Field Theory with Torsion}
\label{sec:2}
Under the extreme conditions following the Big Bang, the universe experienced extremely high 
temperatures and energy densities, where traditional physical theories might fail. Renormalization 
techniques can help adjust these theories to remain effective under high-energy conditions 
using renormalization group theory\,\cite{Bahamonde2023b}. \\
It is well known that the partial derivative of a scalar field produces a covariant vector, but the partial 
derivative of any tensor field cannot form a tensor. However, by introducing supplementary terms, we can transform 
it into a tensor while maintaining the integrity of the tensor field itself. This transformation is called the 
covariant derivative. For a contravariant tensor $F^\alpha$, we can derive its covariant derivative:
\begin{equation}
   \nabla_\beta F^\alpha = \partial_\beta F^\alpha + \Gamma^\alpha_{\beta\gamma} F^\gamma,
   \label{2}
\end{equation}

where $\varGamma^\alpha_{\beta\gamma}$ is called as the Christoffel symbol, 
which ensures the transformation maintains tensor propoties. In general relativity, 
we typically assume the Levi-Civita connection, which is torsion-free, to describe spacetime curvature. 
However, when considering extensions of general relativity or more exotic theories like EC theory
\cite{GarciadeAndrade2022,GarciadeAndrade2024}, torsion can indeed be non-zero and interact with matter 
fields, particularly involving spin. These interactions can indeed lead to significant deviations from 
general relativity under extreme conditions, like near singularities or in highly dense regions.

Let us start our discussion from the classical theory of torsion. In this model, we consider a complex 
scalar field  $\phi$ that is minimally coupled with gravity\,\cite{Hehl1976}.
The Lagrangian $\mathcal{L}$ includes a canonical kinetic term  and a potential $V(\phi)$ of self-interaction 
of $\phi$. The starting action $S_{0}$ is then given by
\begin{align}
S_{0} &= \int d^4 x \sqrt{-g} \mathcal{L} \nonumber \\
      &= \int d^4 x \sqrt{-g} \left( \frac{1}{2} g^{\mu \nu} \nabla_{\mu}\phi^\dagger 
\nabla_{\nu} \phi - V(\phi) \right),
\label{3}
\end{align}

where $\sqrt{-g}$ is the square of negative determinant of the metric tensor $g_{\mu\nu}$, 
ensuring the action is invariant under coordinate transformations, $\nabla_{\mu}$ represents the covariant 
derivative, $\phi^\dagger$ is the Hermitian conjugate of $phi$.

Taking advantage of the unique properties of torsion to facilitate further renormalization efforts, 
an additional non-minimal coupling function term $\xi_{i}$ term is introduced, and torsion is intertwined 
with action $S_{0}$ in a more complex scalar field. To simplify the interaction between torsion and fields, 
we construct new torsion-related tensors $P_{i}, i=1,2,3,4,5$:
\begin{eqnarray}
&& P_1 = R, \quad P_2 = \nabla_\alpha T^\alpha, \quad P_3 = T^\alpha T_\alpha,\nonumber\\
&& P_4 = S^\alpha S_\alpha, \quad P_5 = q^{\alpha\beta\gamma}q_{\alpha\beta\gamma},
\label{4}
\end{eqnarray}
where $P_1$ is the Ricci scalar, representing scalar curvature. $P_{2}$ denotes the divergence of 
the torsion tensor $T^\alpha$, $P_3$ is the norm of the torsion vector $T$, given by $T_\alpha T^\alpha=
g_{\alpha \beta}T^\alpha T^\beta$ with two compents of $T^\alpha$ and $T^\beta$. This term gives a scalar 
quantity representing the magnitude of the torsion tensor in a given spacetime. $P_4$ denoted 
the norm of another tensor field $S^\alpha$. $P_5$ denoted the norm of a rank-3 tensor field 
$q^{\alpha\beta\gamma}$, with the contraction $q^{\alpha\beta\gamma}q_{\alpha\beta\gamma}=
g_{\alpha \alpha^{'}}g_{\beta \beta^{'}}g_{\gamma \gamma^{'}}q^{\alpha\beta\gamma}q^{\alpha{'} 
\beta{'}\gamma{'}}$. This contraction process yields a scalar quantity representing the magnitude of 
$q^{\alpha\beta\gamma}$. 

To describe how the scalar field $\phi$ interacts with gravity 
and torsion within a renormalizable quantum field theory framework, we combine equations 
(\ref{2}),(\ref{3}) with (\ref{4}) to obtain 
\begin{align}
S_{0} &= \int d^4 x \sqrt{-g} \left( \frac{1}{2} g^{\mu\nu} \nabla_\mu \phi^{\dagger} \nabla_\nu \phi \right. \nonumber\\
      &\quad + \frac{1}{2}m^2\phi^2 + \frac{1}{2} \sum_{i=1}^{5} \xi_i P_i \phi^2 \Big),
\label{5}
\end{align}
where $\frac{1}{2}m^2\phi^2$ represents the mass term for the scalar field $\phi$. $\frac{1}{2}
\sum_{i=1}^{5} \xi_{i} P_{i}\phi^2$ includes a sum over five different terms, each representing an 
interaction between the scalar field $\phi$ and different torsion-related quantities $P_i$. The 
scalar field squared $\phi^2$ indicates that these interaction terms are quadratic in the field. 
These terms collectively aim to refine classical torsion models into a renormalizable quantum field 
theory for more precise analysis and predictions.

Given that torsion in the classical spacetime context lacks quantum corrections and leads to consistency, 
making it unrenormalizable, we refine it to a torsion-centered renormalizable quantum field theory 
for more precise analysis and predictions. For our purposes, we start considering the basic renormalizable 
gauge theories in flat spacetime, such as the SM or other Grand Unified Theories (GUTs), and then extend 
these theories to curved spacetime with non-zero torsion. By constructing spinor, vector, 
and scalar fields that include gauge, Yukawa potential\cite{Yukawa1935}, and quartic scalar 
interactions\cite{PeskinSchroeder1995}, we maintain gauge invariance while introducing 
non-minimal interactions between torsion and matter fields. Ultimately, this approach 
yields a renormalizable quantum field theory that includes torsion\,\cite{Shapiro2002}. 
In this framework, the dimensionless coupling of matter self-interactions is not affected 
by torsion, so we only need to consider the interaction between torsion and fields. 
By introducing gauge field, Yukawa interation, and quartic scalar interactions into the 
quantum field theory, we expand equation (\ref{5}) and obtain the following expression
\begin{align}
S_{0} &= \int d^4 x \sqrt{-g} \left[ -\frac{1}{4} G^\alpha_{\beta\gamma}
+ \frac{1}{2} g^{\mu\nu} D_\mu \phi^{\dagger} D_\nu \phi \right. \nonumber\\
&\quad + \frac{1}{2} \left( \sum_i \xi_i P_i + M^2 \right) \phi^2 - V_{\text{int}}(\phi) \nonumber\\
&\quad \left. + i \overline{\psi} \left( \gamma^\alpha D_\alpha + \sum_j \eta_j Q_j - im + h\phi \right) \psi \right] \nonumber\\
&\quad - S_{\text{vac}},
\label{6}
\end{align}

where $G^\alpha_{\beta\gamma}$ denotes the field strength tensor of gauge field containing 
the curvature and dynamics of the field. $D$ is the covariant derivative acting on 
both gravitational and gauge fields, excluding torsion, $M$ is the scale of the GUT symmetry breaking, 
$h$ is a coupling coulping constant determining the strength of these Yukawa interations. 
$V_{\text{int}}(\phi)$ describes the quartic scalar interactions, $\gamma^\alpha (\alpha=0, 1, 2, 3)$ 
denotes the 4$\times$4 Dirac matrices, $\overline{\psi}=\psi^{\dagger}\gamma^{0}$ is the canoical conjugate 
field to the Dirac field $\psi$. The last term, $S_{\rm vac}$, is the vacuum 
action, which is used to maintain renormalizability. Using the above action, we 
can further calculate the effective potential in spacetime with torsion, 
including symmetry breaking and phase transitions caused by curvature and torsion.

In order to simplify the action function, we need to specify the mass and charge of the fields. 
In this paper, we select a charged scalar field $\phi$ and an $\mathrm{SU}(2)$ gauge model with two sets 
of fermion families as an example, then obtain a renormalizable theory in curved 
spacetime in terms of torsion. Let's construct a four-dimensional covariant derivative 
of $\phi$\,\cite{Dimopoulos2002}:
\begin{eqnarray}
 D_{\mu}=\partial_\mu \phi -\frac{ig}{2} \tau^a A^a_\mu \phi,
\label{7}
\end{eqnarray}
where $\tau^a$ are $\mathrm{SU}(2)$ generators, $A^a_\mu$ are the components of $\mathrm{SU}(2)$ 
gauge field, and $g$ is the gauge coupling constant. Combining with equation (\ref{6}), we obtain the Lagrangian,
\begin{eqnarray}
\mathcal{L} &=& -\frac{1}{4} G^a_{\mu\nu} G^{a\mu\nu} 
+ g^{\mu\nu}\left(\partial_\mu \phi^\dagger + \frac{ig}{2} \tau^a A^a_\mu \phi^\dagger \right) \nonumber\\
&& \times \left(\partial_\nu \phi - \frac{ig}{2} \tau^a A^a_\nu \phi \right) 
- \frac{f}{8} (\phi^\dagger \phi)^2 \nonumber\\
&& + \sum_{i=2}^{5} \xi_i P_i \phi^\dagger \phi 
+ i \xi_0 T^\alpha(\phi^\dagger \partial_\alpha \phi - \partial_\alpha \phi^\dagger \phi) \nonumber\\
&& + \sum_{k=1}^{m} i \overline{\chi}^{(k)} (\gamma^\mu \nabla_\mu + 
\sum_{j=1,2} \delta_j Q_j) \chi^{(k)} \nonumber\\
&& + \sum_{k=1}^{m+n} i \overline{\psi}^{(k)}\left(\gamma^\mu \nabla_\mu 
- \frac{ig}{2} \tau^a \gamma^\mu A^a_\mu + \sum_{j=1,2} \mu_j Q_j \right) \psi^{(k)} \nonumber\\
&& - h \sum_{k=1}^{m}(\overline{\psi}^{(k)} \chi^{(k)} \phi 
+ \phi^\dagger \overline{\chi}^{(k)} \psi^{(k)}).
\label{8}
\end{eqnarray}

Due to space constraints, the detailed derivation of the above formula is omitted, but the 
physical meaning of each term and related parameters in equation\,(\ref{8}) must be clearly described as follows:
\begin{enumerate}
\item The term $-\frac{1}{4}G^a_{\mu\nu} G^{a\mu\nu}$ comes from the Yang-Mills Lagrangian, which describes the 
behavior of gauge fields in non-Abelian gauge theories like $\mathrm{SU}(2)$, where $-\frac{1}{4}$ is a normalization factor that 
ensures the correct dimensions and normalization of the kinetic term. $G^a_{\mu\nu} G^{a\mu\nu}$ represents the 
contraction of the field strength tensor with itself, summing over both the space-time indices $\mu,\nu$
 and the internal gauge indices $a$. It gives a scalar quantity that represents the kinetic energy density of 
the gauge fields.
\item The term $g^{\mu\nu}(\partial_\mu \phi^\dagger + \frac{ig}{2} \tau^a A^a_\mu \phi^\dagger)
    (\partial_\nu \phi - \frac{ig}{2} \tau^a A^a_\nu \phi)$ represents the gauge-invariant kinetic energy 
    term of the scalar field $\phi$ in the presence of the gauge field $A^a_\mu$ within a gauge theory framework, 
    specifically for the $\mathrm{SU}(2)$ gauge group.
\item The term $- \frac{f}{8} (\phi^\dagger \phi)^2$ represents a quartic (fourth power) 
      self-interaction involving pairs of $\phi$ field. The negative sign indicates that this term reduces 
     the potential energy of the system, and the $f$ represents the coupling constant associated with the quartic 
    self-interaction term of $\phi$. 
\item  The term $\sum_{i=2}^{5} \xi_i P_i \phi^\dagger \phi$ represents the interation between the scalar field $\phi$
       and various torsion-related tensor $P_i$.
\item The term $i \xi_0 T^\alpha(\phi^\dagger \partial_\alpha \phi - \partial_\alpha \phi^\dagger \phi)$ represents 
      the interaction between the scalar field $\phi$ and the torsion field $T^\alpha$ modulated by the coupling 
      constant $\xi_0$. This term represents the dynamics of the scalar field in the presence of torsion, with the 
      difference $\phi^\dagger \partial_\alpha \phi-\partial_\alpha \phi^\dagger \phi$ ensuring that the interaction 
      respects certain symmetries such as gauge invarience.  
\item The term $\sum_{k=1}^{m} i \overline{\chi}^{(k)}(\gamma^\mu \nabla_\mu + \sum_{j=1,2} 
     \delta_j Q_j) \chi^{(k)}$ represents the kinetic and interaction terms for a set of fermion 
     fields $\chi^{(k)}$ with summation from $k=1$ to $m$. This term captures how $\chi^{(k)}$ move and 
    interact with each other and with other fields, ensuring the theory respects both relativistic and gauge 
    invariance priciples. $\sum_{j=1,2}\delta_j Q_j $ represents additional interaction terms involving the fields $Q_j$, 
   where $\delta_j$ are the coulping constants.
\item The term $\sum_{k=1}^{m+n} i \overline{\psi}^{(k)}(\gamma^\mu \nabla_\mu - \frac{ig}{2} 
       \tau^a \gamma^\mu A^a_\mu + \sum_{j=1,2}\\ \mu_j Q_j) \psi^{(k)}$ represents the kinetic and interaction 
      terms for a set of fermion fields $\psi^{(k)}$ with summation from $k=1$ to $m+n$.
\item The term $-h \sum_{k=1}^{m} (\overline{\psi}^{(k)}\chi^{(k)} \phi + \phi^\dagger 
     \overline{\chi}^{(k)} \psi^{(k)})$ represents Yukawa-type interactions between the scalar field 
   and ferion fields, which are essential for giving mass to the fermions through the Higgs.
\end{enumerate}
In our model, the contribution of fermions is treated as minor, so we will avoid analyzing fermions 
in subsequent calculations. Now, let us examine the renormalizable field theory without Gauss-Bonnet 
correction, whose Lagrangian expression is given as 
\begin{eqnarray}
\mathcal{L} &&= -\frac{1}{4}G^a_{\mu\nu}G^{a\mu\nu} + g^{\mu\nu}\left(\partial_\mu\phi^\dagger
 + \frac{ig}{2}\tau^a A^a_\mu\phi^\dagger \right)\nonumber\\
&&\times\left(\partial_\mu\phi - \frac{ig}{2}\tau^aA^a_\mu\phi\right) - \frac{f}{8}(\phi^\dagger\phi)^2 \nonumber\\
&& + \sum_{i=2}^{5}\xi_i P_i\phi^\dagger\phi  + i\xi_0T^\alpha(\phi^\dagger\partial_\alpha\phi 
   - \partial_\alpha\phi^\dagger·\phi) \nonumber\\
&& \equiv \mathcal{L}_{0} + \triangle\mathcal{L},
\label{9}
\end{eqnarray}
where $\mathcal{L}_{0}$ denotes the Lagrangian of a normal field, and $\Delta \mathcal{L}$ represents 
the correction term to the standard Lagrangian $\mathcal{L}_{0}$, including the following contributions:
\begin{eqnarray}
 \triangle\mathcal{L}=&&-\frac{f}{8}(\phi^\dagger\phi)^2 +\sum_{i=2}^{5}\xi_i P_i \phi^\dagger\phi \nonumber\\
 &&+ i \xi_0 T^\alpha (\phi^\dagger \partial_\alpha \phi-\partial_\alpha \phi^\dagger \cdot \phi).
\label{10}
\end{eqnarray}
We separate out $\Delta \mathcal{L}$ in order to obtain $\Delta S$. The correction term 
in the action, $\Delta S$, helps us determine how new interactions modify the evolution of fields 
and their symmetries. By analyzing $\Delta S$, we can understand the influence of these modifications 
on the physical properties of the system, including potential symmetry breaking, particle interactions, 
and the resulting dynamics. This approach allows us to predict observable consequences of 
the new interactions introduced by $\Delta \mathcal{L}$.

The variation of the action $\Delta S$ is then obtained in the following form
\begin{eqnarray}
\Delta S&&=\int d^4 x \delta(\sqrt{-g}\triangle\mathcal{L})\nonumber\\
   &&=\delta \int (-\frac{1}{8}f)(\phi^\dagger \phi)^2\sqrt{-g}d^4 x  \nonumber\\
  &&+ \delta \int \xi_2 \frac{1}
{\sqrt{-g}}\partial_\mu(\sqrt{-g}T^\mu)\sqrt{-g}\phi^\dagger\phi d^4 x \nonumber\\
&&+ \delta\int\xi_3 T_\alpha g^{\alpha\beta}T_\beta\sqrt{-g}\phi^\dagger\phi d^4x \nonumber\\
 &&+\delta\int\xi_4 S_\alpha g^{\alpha \beta}S_\beta\phi^\dagger\phi\sqrt{-g}d^4 x \nonumber\\
&&+ \delta\int\xi_5\left(q^{\alpha\beta\gamma}g^{\alpha \alpha^{'}}g^{\beta\beta^{'}}g^{\gamma\gamma^{'}}q_{\aleph^{'}
\beta^{'}\gamma^{'}}\right)\sqrt{-g}\phi^\dagger\phi d^4x   \nonumber\\
 && +\delta\int i\xi_0T^\alpha(\phi^\dagger\partial_\alpha\phi-\partial_\alpha\phi^\dagger\phi)\sqrt{-g}d^4x.
\label{11}
\end{eqnarray}
By utilizing $\Delta S$, we obtain the modified energy-momentum tensor, essential for deriving the 
Einstein field equations with the Gauss-Bonnet correction term. This approach not only provides corrections to the 
gravitational field equations but also informs conservation laws and other dynamical equations that govern the 
evolution of fields in modified gravity theories. Additionally, $\Delta \mathcal{L}$ facilitates 
direct computation of significant results, including the modified Einstein field equations, which are 
crucial for understanding the impact of higher-order curvature corrections on spacetime geometry. For further 
details, refer to the next section.
\section{New Einstein's equations with GB Term}
\label{sec:3} 
As we know, the Einstein field equations (EFE), also known as Einstein's equations, are a set of 
ten interrelated differential equations in Albert Einstein's theory of general relativity. 
These equations describe how matter and energy in the universe influence the curvature of spacetime.
As a significant source term of the gravitational field, the stress-energy tensor, which represents 
the distribution of matter and energy, directly encapsulates the gravitational field's properties 
and the influence of matter fields on the geometric structure of spacetime. 

In order to  obtain the energy-momentum tensor within the Gauss-Bonnet 
theory, it's necessary to integrate the Gauss-Bonnet term into the action. 
For the purpose of exploring  the potential of the Gauss-Bonnet theory, we extend 
the aforementioned calculations to the Einstein's equations, considering the action 
with Gauss-Bonnet gravity 
	\begin{equation}
	S=\int d^4x\sqrt{-g}\left[\frac{R}{2\kappa}+f(\mathcal{G})\right]+S_M(g_{\mu\nu},T,\phi),
          \label{12}
	\end{equation}
where $R/(2\kappa)$ is the Einstein-Hilbert term, representing the curvature of spacetime, 
$\kappa = 8\pi G$, with $G$ being the gravitational constant, and $S_M(g_{\mu\nu}, T, \phi)$ is the action 
of matter fields, depending on the metric tensor $g_{\mu\nu}$, torsion $T$, and scalar field $\phi$. 
By varying the given action with respect to the metric tensor, one obtains Einstein's 
equations with the Gauss-Bonnet correction term,
\begin{eqnarray}
G_{\mu\nu} &&+ 8 [R_{\mu\rho\nu\sigma} + R_{\rho\nu}g_{\sigma\mu} - R_{\rho\sigma}g_{\nu\mu} 
 -R_{\mu\nu}g_{\sigma\rho} \nonumber\\
&&  + R_{\mu\sigma}g_{\nu\rho} + \frac{R}{2}(g_{\mu\nu}g_{\sigma\rho}-g_{\mu\sigma}g_{\nu\rho})]\nonumber\\ 
&&\times\nabla^\rho\nabla^\sigma \dot{f}+ (\mathcal{G}\dot{f}- f)g_{\mu\nu} = \kappa T_{\mu\nu},
\label{13}
\end{eqnarray}
where $G_{\mu\nu}=R_{\mu,\nu}-\frac{1}{2}Rg_{\mu,\nu}$ is the Einstein tensor, 
the item with the parentheses is the GB correction tensor, $\nabla^\rho\nabla^\sigma \dot{f}$ is 
the second covariant derivatives of $\dot{f}$, and $T_{\mu\nu}$ is 
the energy-momentum tensor, defined as,
	\begin{equation}
		T_{\mu\nu}=-\frac{2}{\sqrt{-g}}\frac{\delta(\sqrt{-g}\mathcal{L}_M)}{\delta g^{\mu\nu}},
            \label{14}
       \end{equation}
where $\mathcal{L}_M$ is the Lagrangian density of the matter fields. To derive a simplified version of 
energy-momentum tensor, we start from a simple scenario, considering 
a spatially flat Friedmann-Lemaître-Robertson-Walker\,(FLRW) universe with a metric,
	\begin{equation}
		ds^2=-dt^2+a^2(t)(dx^2_1+dx^2_2+dx^2_3),
           \label{15}
	\end{equation}
where the scale factor $a(t)$ is a crucial parameter for understanding the dynamics of the universe.
It is closely related to the Hubble parameter $H=a/\dot{a}$, describing the rate of expansion of the universe.
 Inserting equation\,(\ref{15})into equation\,(\ref{13}), we get
	\begin{eqnarray}
T_{00} &&= 3H^2 + (\mathcal{G} \dot{f} - f)(-1) \nonumber \\
       &&+ 24 \ddot{f} \left( \ddot{H} H^5 + 4H^6 \dot{H} + 2 \dot{H}^2 H^4 \right),
\label{16}
\end{eqnarray}
\begin{equation}
T_{0i} = 0, \quad \quad \quad \quad T_{i0} = 0,
\label{17}
\end{equation}
\begin{eqnarray}
T_{ij} &&= -a^2(2 \dot{H} + 3H^2) \delta_{ij} + 8 \left( \ddot{f} \ddot{\mathcal{G}} + \dddot{f} 
\dot{\mathcal{G}}^2 \right) \nonumber \\
&& + a^2 \delta_{ij} (-H^2) - a^2 \ddot{f} \dot{\mathcal{G}} \delta_{ij} 10H^3.
\label{18}
\end{eqnarray}
Here the first term in equation\,(\ref{18}) shows the influence of the Hubble parameter $H$
and its time derivative $\dot{H}$ on spatial stress, describing standard cosmic expansion dynamics 
and linking to the Friedmann equation, the second term includes the Gauss-Bonnet invariant $\mathcal{G}$ 
and its derivatives due to the dynamic Universe and spacetime, indicating higher-order gravitational 
corrections, and the third term highlights the complex influence of Gauss-Bonnet corrections on the Hubble 
parameter and spacetime geometry, showing how time variations of $\mathcal{G}$ affect cosmic expansion.
The terms $\ddot{f}$ and $\dddot{f}$ reflect the effects of spacetime curvature, especially during rapid 
Hubble rate changes, modifying standard gravity theories and aiding in understanding dark energy and inflation
\cite{koh2014inflation}.

We want to apply the renormalizable field theory to correct the energy-momentum tensor. This 
approach allows higher-order interactions and corrections, which might introduce divergent energy or 
momentum, to be eliminated through renormalization, ensuring the theory remains consistent even at 
high energy limits. The previous defined $\Delta L$ in equation\,(\ref{10}) is the part we should consider modifying.
After simplification, we can obtain the contribution of $\triangle\mathcal{L}$ to the energy-momentum tensor,
\begin{eqnarray}
\triangle T_{\mu\nu} &=& -\frac{2}{\sqrt{-g}} \frac{\delta(\sqrt{-g} \triangle \mathcal{L})}{\delta g^{\mu\nu}} \nonumber\\
&=& -2 \Bigg\{ \left[ -\frac{1}{8} f(\phi^\dagger \phi)^2 + \xi_5 q_{\alpha \beta \gamma} q^{\alpha \beta \gamma} \right]
\left( -\frac{1}{2} g_{\mu\nu} \right) \nonumber\\
&& + \left[ -\xi_2 T_\mu^{\prime} \partial_\nu^{\prime} (\phi^\dagger \phi) + \xi_3 T_{\mu^{\prime} \nu^{\prime}} \phi^\dagger \phi
+ \xi_4 S_\mu^{\prime} S_\nu^{\prime} \phi^\dagger \phi \right] \nonumber\\
&& + \xi_5 \left( q_{\mu^{\prime} \beta \gamma} q_{\nu^{\prime}}^{\beta \gamma} + q_{\alpha \mu^{\prime} \gamma} q^{\alpha \gamma}_{\nu^{\prime}} 
+ q_{\alpha \beta \mu^{\prime}} q^{\alpha \beta}_{\nu^{\prime}} \right) \nonumber\\
&& \left. + i\xi_0 \text{Tr}\left( \phi^\dagger \partial_\alpha \phi - \partial_\alpha \phi^\dagger \phi \right) \right] \nonumber\\
&& \times \left[ g^{\mu' \nu'} \left( -\frac{1}{2} g_{\mu\nu} + \delta^{\mu'}_\mu \delta^{\nu'}_\nu \right) \right] \Bigg\},
\label{19}
\end{eqnarray}

where the Kronecker delta parts $\delta^{\mu^{'}}_\mu\delta^{\nu^{'}}_\nu$ ensure the correct transformation 
properties of the indices, $\phi^\dagger\phi$ is the trace of the matrix $\Phi^\dagger\Phi$, and $\Phi$ is a matrix 
representation of $\phi$,
        \begin{equation}
	\phi^\dagger \phi \equiv \mathrm{Tr}(\Phi^\dagger \Phi) = \phi_0^2.
       \label{20}
      \end{equation}
In order to facilitate calculations and derivations, we consider the simplest matrix 
 representation $\Phi$,
	     \begin{equation}
		\Phi=\frac{\phi_0}{\sqrt{2}}\left[{\begin{array}{ccccc}
				0&0&0&0&0\\
				0&0&0&0&0\\
				0&0&0&0&0\\
				0&0&0&0&1\\
				0&0&0&-1&0\\
		\end{array}}\right].
        \label{21}
	\end{equation}
Taking variations of $S_\mu$ and $q_{\alpha\beta\gamma}$ \cite{ToriiShinkai2008}, we obtain,
\begin{eqnarray}
\Delta T_{\mu \nu} &=& -\frac{1}{4} g_{\mu\nu} g^{\mu' \rho'} g^{\nu' \sigma'} 
\left( G_{\mu' \nu'}^a G_{\rho' \sigma'}^a + G_{\mu' \nu'}^0 G_{\rho' \sigma'}^0 \right) \nonumber \\
&& + g^{\nu' \sigma'} \left( G_{\mu \nu'}^a G_{\nu \sigma'}^a + G_{\mu \nu'}^0 G_{\nu \sigma'}^0 \right) 
- \frac{1}{8} f g_{\mu\nu} \phi_0^2 \nonumber \\
&& + \left[ 2\xi_2 T_{\mu'} \partial_{\nu'} \phi_0^2 - (D_{\mu'} \Phi)^\dagger_{ji} (D_{\nu'} \Phi)_{ij} \right. \nonumber \\
&& \left. - 2\xi_3 T_{\mu'} T_{\nu'} \phi_0^2 \right] \nonumber \\
&& \times \left( -\frac{1}{2} g_{\mu\nu} g^{\mu' \nu'} + \delta^{\mu'}_\mu \delta^{\nu'}_\nu \right).
\label{22}
\end{eqnarray}

To establish the correlation between the energy-\\momentum tensor and torsion in Gauss-Bonnet theory, we 
need to consider the previously derived general energy-momentum tensor and modify it with the Gauss-Bonnet 
term. Here are the improved field equations and their components:

(1) The Energy Density Component:
\begin{eqnarray}
T_{00} &&= \frac{1}{4} g^{\mu' \rho'} g^{\nu' \sigma'} \left[ \text{Tr} \left( G_{\mu' \nu'} G_{\rho' \sigma'} \right) + G^0_{\mu' \nu'} G^0_{\rho' \sigma'} \right] \nonumber \\
&&- \frac{1}{2} g^{\rho' \nu'} \left[ \text{Tr} \left( D_{\rho'} \Phi D_{\nu'} \Phi \right) 
+ D_{\eta'} S D_{\nu'} S \right] \nonumber \\
&&+ g^{\nu' \sigma'} \left[ \text{Tr} \left( G_{0 \nu'} G_{0 \sigma'} \right) + G^0_{0 \nu'} G^0_{0 \sigma'} \right] \nonumber \\
&&- \text{Tr} \left[ (D_0 \Phi)^\dagger D_0 \Phi + \nabla_0 S \nabla_0 S \right] \nonumber \\
&&+ \left\{ \frac{1}{4} \lambda \left[ \text{Tr} \left( \Phi^\dagger \Phi \right) - M^2 \right]^2 \right\} \nonumber \\
&&+ \frac{1}{2} h S^2 \text{Tr} \left( \Phi^\dagger \Phi \right) + V_S(S),
\label{23}
\end{eqnarray}
where $G_{\mu'\nu'}$ and $G_{\rho'\sigma'}$ are the gauge fields, and $G^0_{\mu'\nu'}$ and $G^0_{\rho'\sigma'}$ are 
their temporal components. $D_{\eta'}SD_{\nu'}S$ is the product of derivatives of the torsion field $S$, $\nabla_{0}
S\nabla_{0}S]$ is the product of the covariant derivative of the field $S$ with itself, describing how the torsion 
field varies with respect to time and the interaction between these variations, $M^2$ is the mass term, and 
$V_S(S)$ is potential of the torsion field.

(2)The flux of energy in the $i$-th spatial direaction:
   \begin{eqnarray}
	T_{0i}&&=0=g^{\nu'\sigma'}[Tr(G_{0\nu'}G_{i\sigma'})+G_{0\nu'}^0G_{i\sigma'}^0]\nonumber\\
 &&-[Tr((D_0\Phi)^{\dagger}D_i\Phi)+\nabla_0S\nabla_iS],
    \label{24}
   \end{eqnarray}
where $Tr(G_{0\nu'}G_{i\sigma'})$ is the trace of the product of the gauge field components
 $G_{0\nu'}$ and $G_{i\sigma'}$, $Tr((D_0\Phi)^{\dagger}D_i\Phi)$ is the trace of the conjugate of 
the covariant derivative of the scalar field $\Phi$ with respect to the temporal component and its 
covariant derivative with respect to the spatial component, $\nabla_0S\nabla_iS$ is the product of 
the covariant derivative of $S$ with respect to the time component and the spatial component, describing 
the variation of the torsion field in time and space and its influence on the overall dynamics. 

(3)The flux of momentum in the time direction:
   \begin{eqnarray}
	T_{i0}=0&&=g^{\nu'\sigma'}\left[Tr(G_{i\nu'}G_{0\sigma'})+G_{i\nu'}^0G_{0\sigma'}^0 \right]\nonumber\\
 &&-\left[Tr((D_i\Phi)^{\dagger}D_0\Phi)+\nabla_0S\nabla_iS \right],
 \label{25}
   \end{eqnarray}
where $Tr(G_{i\nu'}G_{0\sigma'})$ is the trace of the product of the gauge field components $G_{i\nu'}$ 
and $G_{0\sigma'}$, $Tr((D_i\Phi)^{\dagger}D_0\Phi)$ is the trace of the conjugate of the covariant 
derivative of the scalar field $\Phi$ with respect to the spatial component and its covariant derivative 
with respect to the temporal component. In equations\,(\ref{24}) and (\ref{25}), $T_{i0}=0$ and $T_{i0}=0$ 
show the balance of gauge, scalar, and torsion fields ensuring no net energy or momentum flux in the spatial 
direction, satisfying the specified conditions.
 
(4)The spatial stress component:
\begin{eqnarray}
T_{ij}=&&-\frac{1}{4} \delta_{ij} a^2 g^{\mu' \rho'} g^{\nu' \sigma'} \left[ \text{Tr}(G_{\mu' \nu'} G_{\rho' \sigma'}) 
+ G_{\mu' \nu'}^0 G_{\rho' \sigma'}^0 \right] \nonumber \\
&&+ g^{\nu' \sigma'} \left[ \text{Tr}(G_{i \nu'} G_{0 \sigma'}) + G_{i \nu'}^0 G_{j \sigma'}^0 \right] \nonumber \\
&&- \left[\text{Tr}(D_i \Phi)^\dagger D_j \Phi + \nabla_i S \nabla_j S \right] \frac{1}{2} \delta_{ij} a^2 g^{\mu' \nu'} \nonumber \\
&&\times \left[ \text{Tr}(D_{\mu'} \Phi)^\dagger D_{\nu'} \Phi + \nabla_{\mu'} S \nabla_{\nu'} S \right] \nonumber \\
&& - a^2 \delta_{ij} \left\{ \frac{1}{4} \lambda \left[ \text{Tr}(\Phi^\dagger \Phi) - M^2 \right]^2 \right. \nonumber \\
&& \left. + \frac{1}{2} h S^2 \text{Tr}(\Phi^\dagger \Phi) + V_S(S)\right \},
\label{26}
\end{eqnarray}
where $\delta_{ij}$ is the Kronecker delta function, ensuring the correct indexing of spatial components. 
$M^2$ is the mass term, $V_S(S)$ is the potential energy of the torsion field $S$. This equation is composed 
of four terms: the first term represents the contribution to the spatial stress tensor from the gauge fields 
$G$ and their temporal components $G^0$, the second term represents additional contributions from the gauge 
fields and their interactions between different components, the third term represents the contribution from 
the scalar field $\Phi$ and the torsion field $S$, the fourth term represents the contribution from the 
potential energy terms of $\Phi$ and $S$. 

The energy-momentum tensor in this theoretical framework reflects the attributes of renormalizable matter 
fields. Gauge fields contribute through their kinetic terms and interactions between spatial and temporal 
components, affecting the overall stress-energy distribution. Scalar fields $\Phi$ and $S$ influence the 
tensor through their kinetic energies and potential terms, which include self-interactions and couplings 
between fields. These terms illustrate how the dynamics of gauge and scalar fields impact spacetime 
geometry, particularly in scenarios such as cosmological evolution or gravitational interactions. The 
Gauss-Bonnet correction enhances efficiency in environments with extreme gravitational 
forces\,\cite{Dutta2018,Fernandes2022b}. This energy-momentum tensor provides insight into observability 
and other pertinent characteristics.
\section{Summary and Discussion}
\label{sec:5}
In this paper, we aim to establish a renormalizable field within the Gauss-Bonnet theory, setting the 
foundation for future research. According to \cite{Bhattacharjee2017}, the Gauss-Bonnet term is topological 
in four dimensions, but when coupled with fields, it induces dynamical effects that influence the evolution 
of these fields. This makes the Gauss-Bonnet theory uniquely advantageous for studying the early, 
high-energy universe. We considered the common $\mathrm{SU}(2)$ gauge model. 

By extending the standard theory to a flat universe, we explored scenarios with non-zero torsion. 
We then rewrote the action using the torsion's unique spacetime characteristics to achieve field renormalization 
and derived the field's energy-momentum tensor. This tensor directly determines the field's gravitational 
properties and dynamics, impacting the Hubble constant and enabling the exploration of Gauss-Bonnet theory 
from an astronomical perspective. Using the modified energy-momentum tensor, we revised the field equations 
of the supersymmetric hybrid inflation model, reflecting the evolutionary properties of the field.

In future work, we aim to investigate more specific models to derive clearer gravitational characteristic 
equations, thus simplifying the Gauss-Bonnet theory and identifying models highly sensitive to parameters. 
By extremizing or adding constraints to our generalized field equations, we can better calculate possible 
states of the early universe, providing a theoretical basis for observations. Currently, our theory lacks 
specific model calculations, which may lead to new challenges in subsequent computations requiring further 
research and exploration.

\begin{acknowledgements}
This work was supported by the National Key Research and Development Program of China (Grant Nos. 2021YFC2203501, 
2021YFC2203502, 2021YFC2203503, and 2021YFC2203600), the National Natural Science Foundation of China (Grant Nos. 
12173077, 11873082, 11803080, and 12003062), and the Scientific Instrument Developing Project of the Chinese Academy 
of Sciences (Grant No. PTYQ2022YZ\\ZD01).
\end{acknowledgements}

\end{document}